\documentclass[11pt,a4paper]{article}
\usepackage{amsmath,amssymb}
\usepackage{color}
\usepackage{graphicx}
\usepackage{subfig}
\usepackage{cite}                %%是引用格式变漂亮  %%PRD等APS模板必须禁用该项
\usepackage{hyperref}            %%超链接必须
\usepackage{multirow,makecell}
\usepackage{braket,bm}
\usepackage{physics}
\usepackage{cancel}
\RequirePackage{doi}
\usepackage{hyperref}
\usepackage{leftidx}
\usepackage[text={17cm,24.5cm},centering]{geometry}
\usepackage{textcomp}
\usepackage{wasysym}
\usepackage{mathtools}
\numberwithin{equation}{section}

\def \be {\begin{equation}}
\def \ee {\end{equation}}
\def \ba {\begin{array}}
\def \ea {\end{array}}
\def \bea{\begin{eqnarray}}
\def \eea{\end{eqnarray}}

\def \a {\alpha}
\def \b {\beta}
\def \g {\gamma}

\def \d {\delta}
\def \D {\Delta}
\def \dg {\dagger}

\def \s {\sigma}

\def \mF {\mathcal F}

\def \mH {\mathcal H}
\def \mI {\mathcal I}

\def \mO {\mathcal O}

\def \tr {\textrm{tr}}

\def \and {{\textrm{and}}}

\begin{document}
\begin{titlepage}
	
	\title{\textbf {Entanglement asymmetry in the Hayden-Preskill protocol}}
	\author{Hui-Huang Chen$^{a, b}$\footnote{chenhh@jxnu.edu.cn}~, Zi-Jun Tang$^{a}$}
	\date{}
	
	\maketitle
	\underline{}
	\vspace{-12mm}
	
	\begin{center}
		{\it
            $^{a}$ College of Physics and Communication Electronics, Jiangxi Normal University,\\ Nanchang 330022, China\\ 
            $^{b}$ SISSA and INFN Sezione di Trieste, via Bonomea 265, 34136 Trieste, Italy
             
		}
		\vspace{10mm}
	\end{center}
	\begin{abstract}
	 In this paper, we consider the time evolution of entanglement asymmetry of the black hole radiation in the Hayden-Preskill thought experiment. We assume the black hole is initially in a mixed state since it is entangled with the early radiation. Alice throws a diary maximally entangled with a reference system into the black hole. After the black hole has absorbed the diary, Bob tries to recover the information that Alice thought should be destroyed by the black hole. In this protocol, we found that a $U(1)$ symmetry of the radiation emerges before a certain transition time (the time when the vanishing entanglement asymmetry begins to grow). This emergent symmetry is exact in the thermodynamic limit and can be characterized by the vanishing entanglement asymmetry of the radiation. The transition time depends on the initial entropy and the size of the diary. What's more, when the initial state of the black hole is maximally mixed, this emergent symmetry survives during the whole procedure of the black hole radiation. We successfully explained this novel phenomenon using the decoupling inequality.
	\end{abstract}
	
\end{titlepage}

\thispagestyle{empty}

\newpage

\tableofcontents
\section{Introduction}
The black hole information paradox, first raised by Stephen Hawking \cite{Hawking:1975vcx, Hawking:1976ra}, poses a significant challenge to our understanding of fundamental physics. Hawking's work suggests that black holes emit radiation (known as Hawking radiation) and eventually evaporate, leading to the loss of information about the matter that initially fell into the black hole. This appears to contradict the principles of quantum mechanics, which assert that information should be conserved. For recent reviews, see \cite{Harlow:2014yka, Almheiri:2020cfm}.

\par Don Page first suggested that the time evolution of a black hole over a sufficiently long period (longer than the scrambling time) can be described by a Haar-random unitary \cite{Page:1993df, Page:1993wv}. Page's calculations showed that the entropy of the Hawking radiation initially increases, reaches a maximum at the so-called Page time, and then decreases as the black hole continues to evaporate. This behavior is crucial because it implies that information is not lost but is instead encoded in the Hawking radiation. 

\par One of the most notable contributions by Don Page is the concept of Page time. Page time refers to the point during the black hole evaporation process when the black hole has emitted half of its initial entropy in the form of Hawking radiation. Before this point, the radiation is highly mixed and contains little information about the initial state of the black hole. After this point, the radiation starts to carry more information about the initial state, making it possible to reconstruct the original information.

\par
\begin{figure}
        \centering
        \subfloat
        {\includegraphics[width=6cm]{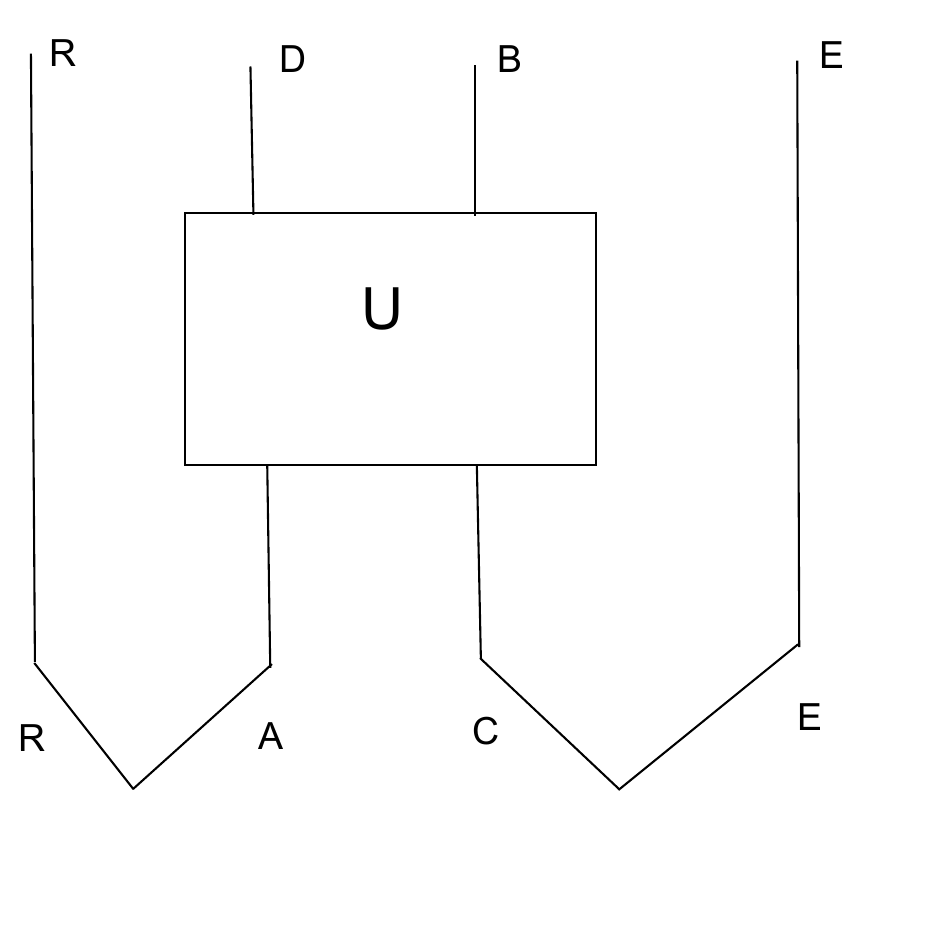}} 
        \caption{The Hayden-Preskill protocol: Alice throws $N_A$ qubits diary $A$, which is maximally entangled with $R$, into the initial black hole $C$. The black hole $C$ is entangled with its early radiation $E$. We only assume that the black hole is in a general mixed state with R\'enyi-two entropy equals to $s$. While in the original paper of Hayden and Preskill, they assume that the black hole was just past the Page time and thus is maximally mixed.}
        \label{fig1}
\end{figure}

 Hayden and Preskill consider a more interesting setup \cite{Hayden:2007cs}: Alice initially prepares the diary maximally entangled with a reference system $R$ and waits until after the Page time to throw the diary $A$ into the black hole $C$. In this scenario, the black hole $C$ is maximally entangled with its early radiation $E$ (as shown in figure \ref{fig1}). The quantum evolution of the black hole can be modeled by applying a Haar-random unitary operator $U$ to the joint system $AC$. Subsequently, the system can be reinterpreted as a tensor product of Hawking radiation $B$ and the remaining black hole $D$. Now, Bob wants to recover the information that Alice has thrown into the black hole. Hayden and Preskill demonstrated that this is possible, and when Bob has access to the early radiation, the information essentially comes out as fast as it possibly could. This is why they refer to "old" black holes as information mirrors. 

\par The Hayden-Preskill protocol highlights the concept of quantum scrambling, where information is rapidly mixed and distributed throughout a quantum system. This is a phenomenon of interest not only in black hole physics but also in many-body quantum systems and quantum computing \cite{Yoshida:2017non, Yoshida:2018vly, Bao:2020zdo, Cheng:2019yib, Piroli:2020dlx, Hayata:2021kcp}. The ability to scramble and unscramble information efficiently is crucial for developing robust quantum error correction codes and quantum algorithms.

\par Recently, entanglement asymmetry was proposed as a quantity to characterize the breaking and restoration of symmetries \cite{Ares:2022koq, Ares:2023kcz}. On one hand, it has been used as a tool to study the quantum version of the Mpemba effect \cite{Murciano:2023qrv}, whose classical counterpart states that hot water can freeze faster than cold water. Since then, entanglement asymmetry and the quantum Mpemba effect have been extensively studied in Hamiltonian dynamics \cite{Murciano:2023qrv, Rylands:2023yzx, Rylands:2024fio, Liu:2024uqf, Ares:2024nkh}, quantum field theories\cite{Chen:2023gql, Fossati:2024ekt, Benini:2024xjv, Kusuki:2024gss}, and random circuits \cite{Bertini:2023ysg, Liu:2024kzv, Turkeshi:2024juo, Foligno:2024jpq}. More recently, the quantum Mpemba effects have been observed on quantum simulation platforms \cite{Joshi:2024sup}. On the other hand, the interplay between entanglement and symmetries in random systems and black hole problems has also attracted much attention in recent years, including studies in\cite{Bianchi:2019stn, Nakata:2020vvy, Murciano:2022lsw, Lau:2022hvc, Ares:2023ggj, Russotto:2024pqg}. In these studies, symmetry-resolved entanglement and entanglement asymmetry play pivotal roles
\par In paper \cite{Murciano:2022lsw, Lau:2022hvc}, the authors examine the case where a Haar-random pure state preserves a $U(1)$ symmetry, and in this situation, the symmetry-resolved entanglement entropy can be defined and the corresponding symmetry-resolved Page curves are derived. In paper \cite{Ares:2023ggj}, the authors investigate when a black hole and the radiation can have emergent symmetries. They use the same qubit model of black hole as in \cite{Murciano:2022lsw, Lau:2022hvc} but adopt the opposite assumption, \textit{i.e.}, they start with a Haar-random pure state that lacks any symmetry. By calculating the average entanglement asymmetry of the emitted radiation, they find that a $U(1)$ symmetry emerges before the Page time. The mixedness of the black hole's initial state, a key feature of the Hayden-Preskill protocol, along with the findings in \cite{Ares:2023ggj}, motivates us to investigate whether emergent symmetries of black holes and radiation can arise when evolving from a mixed state.
\par In this paper, we focus on the entanglement asymmetry in the Hayden-Preskill protocol for a general initial mixed state (not necessarily being maximally mixed as in the original paper of Hayden and Preskill). In the black hole problem, this corresponds to the time evolution of the entanglement asymmetry of radiation. In paper \cite{Nakata:2020vvy}, the authors extend the original model of Hayden-Prskill to the case where the system has symmetries. They investigate the role of symmetries in the process of information recovery. However, in this work, we assume that no symmetries exist initially: neither the initial state of the black hole nor the Haar-random evolution operator preserves any symmetries. In \cite{Li:2021mnl}, the authors consider a similar setup to ours, but focus on the scrambling and decoding problem in the Hayden-Preskill protocol at finite temperatures.  
\par Entanglement asymmetry is defined as
\be 
\D S(\rho_A)=S(\rho_A)-S(\rho_{A,Q}).
\ee
Some explanations about this definition are needed here. $\rho_A=\tr_{\bar A}\rho$ is the reduced density matrix of subsystem $A$ and $S(\rho_A)=-\tr\rho_A\log\rho_A$ is the von Neumann entropy of $\rho_A$. While $\rho_{A,Q}$ is obtained by removing the off-diagonal elements of $\rho_A$
\be 
\rho_{A,Q}=\sum_q\Pi_q\rho_A\Pi_q, 
\ee
where $\Pi_q$ is the projector onto the $q$-th eigenspace of the corresponding symmetry operator. One can define the R\'enyi entanglement asymmetry as
\be
	\Delta S^{(n)}(\rho_A)=S^{(n)}(\rho_{A,Q})-S^{(n)}(\rho_A)=\frac{1}{1-n}\left[\log\mathrm{Tr}(\rho_{A, Q}^n)-\log\mathrm{Tr}(\rho_A^n)\right].
\ee 
Then $\Delta S(\rho_A)$ can be accessed from $\Delta S^{(n)}(\rho_A)$  by taking the limit $\Delta S(\rho_A)=\lim_{n\rightarrow 1}\Delta S^{(n)}(\rho_A)$.

\section{The setup}
\par 
We model the merged system (diary $A$ plus initially mixed black hole $C$) with an $N$-qubit system $S$ whose dynamics is governed by a random unitary $U$. The local basis of each qubit is denoted by $\ket{0}_i,\ket{1}_i$ with $i=1,2,\cdots,N$. The global vacuum $\ket{0}$ is defined as $\ket{0}=\otimes_{i=1}^N\ket{0}_i$. The dimension of the Hilbert space of $S$ is $d=2^{N}$ and $U$ is a $2^N\times 2^N$ random unitary matrix. We also introduce the $U(1)$ symmetry group generator $Q$, defined by $Q=\sum_{i=1}^N\ket{1}_i\bra{1}$. Clearly, $Q$ counts the number of excitations above the vacuum.
\par We initially prepare $d_A=2^{N_A}$ orthonormal states that only differ on the first $N_A$ qubits, 
\be\label{rhoi} 
\rho_i=\ketbra{i_A}{i_A}\otimes\rho   \qquad i=1,2,\cdots,d_A,
\ee
where $\ket{i_A}$ are the orthonormal basis of the Hilbert space of subsystem $A$ and $\rho$ is the state of the remaining $N-N_A$ qubits. We make no assumptions on the explicit form of $\rho$. Therefore, in general $[Q,\rho_i]\neq 0$. We also introduce the canonical purification of $\rho$ as $\ket{\sqrt{\rho}}$, which means $\rho=\tr_E\ketbra{\sqrt{\rho}}{\sqrt{\rho}}$. The pure state $\ket{\sqrt{\rho}}$ lives in the Hilbert space of the system $C$ and the early radiation $E$. Before evolving the system by applying a random unitary, we maximally entangle the first $N_A$ qubits with a reference system $R$ with the same size.

\par The initial state of the whole system (including the reference qubits and the early radiation) can be written as
\be 
\ket{\Psi}=\left(\frac{1}{\sqrt{d_A}}\sum_{i=1}^{d_A}\ket{i_R}\ket{i_A}\right)\otimes\ket{\sqrt{\rho}}.
\ee
As mentioned before, we describe the evolution of the merged system by a Haar random unitary $U$\footnote{We assume that the black hole absorbs the diary instantly.}. Note that the operator $U$ only acts on the $N$-qubit system and does not have effect on the early radiation. Then the time-evolved state is 
\be 
\ket{\Psi(t)}=\frac{1}{\sqrt{d_A}}\sum_{i=1}^{d_A}\ket{i_R}U_S\otimes I_E\ket{i_A,\sqrt{\rho}},
\ee
where we denote the $N$-qubit system as $S$ and $S=A\cup C=B\cup D$.
\section{Averaged purity of reduced state}
\par Tracing out the early radiation $E$ and the reference qubits $R$, we obtain the reduced density matrix $\rho_S$ of the $N$-qubit system as
\be 
\rho_S=\frac{1}{d_A}\sum_{i=1}^{d_A}U\rho_iU^{\dg},
\ee
where $\rho_i$ is defined in eq.~(\ref{rhoi}) and we have omitted the subscript in $U_S$ for simplicity. Here, again, no symmetry restrictions on $U$ are imposed. We have $[\rho_S,Q]\neq 0$ and then $[Q_B,\rho_B]\neq 0$, which allows one to define the entanglement asymmetry of $\rho_B$. 
\par To compute the averaged purity of $\rho_B$ \textit{i.e.} $\mathbb{E}_U[\tr(\rho_B^2)]$, we can first compute $\mathbb{E}_U[\tr(\rho_S^{\otimes2})]$
\be\label{ErhoS2} 
\begin{split}
&\mathbb{E}_U[\rho_S^{\otimes 2}]=\frac{1}{d_A^2}\sum_{i,j=1}^{d_A}\mathbb{E}_U[U\rho_i U^{\dg}\otimes U\rho_j U^{\dg}]. 
\end{split}
\ee
\par For evaluating the expectation value involving four Haar random unitary matrices $U$ and two operators $\mO,\mO'$, the following formula is useful
\be\label{formula} 
\mathbb{E}_U[U\mO U^{\dg}\otimes U\mO' U^{\dg}]=\sum_{\s=\pm}\frac{\tr\mO\tr\mO'+\s~\tr(\mO\mO')}{d(d+\s)}\frac{I+\s\mF}{2},
\ee
where $I$ and $\mF$ are the identity operator and the swap operator on the two ``replicas" of the system $S$, respectively. Denoting $\ket{\a},\ket{\b}$ as the basis of the Hilbert space of $S$, one can write $I=\sum_{\a,\b\in\mH_S}\ketbra{\a}{\a}\otimes\ketbra{\b}{\b}$ and $\mF=\sum_{\a,\b\in\mH_S}\ketbra{\a}{\b}\otimes\ketbra{\b}{\a}$. 
\par Using the above formula, it's convenient to analyze the diagonal terms ($i=j$ terms) and cross terms ($i\neq j$ terms) in eq.~(\ref{ErhoS2}) separately. All the diagonal terms give the same contribution
\be\label{first} 
\begin{split}
&\mathbb{E}_U[U^{\otimes2}\rho_i^{\otimes2}U^{\dg\otimes2}]=\frac{(\tr\rho)^2I+\tr(\rho^2)\mF}{d^2-1}-\frac{(\tr\rho)^2\mF+\tr(\rho^2)I}{d(d^2-1)}\\
&=\frac{I+2^{-s}\mF}{d^2-1}-\frac{\mF+2^{-s}I}{d(d^2-1)}.
\end{split}
\ee
Here we have defined $s\equiv-\log_2\tr(\rho^2)$ as the initial R\'enyi-2 entropy of the black hole. Since all the $\rho_i$ are density matrices, we have $\tr\rho_i=\tr\rho=1$. We have also assumed they are mixed states, so  $\tr\rho_i^2=\tr(\rho^2)\neq 1$. From the definition of $\rho_i$, it's obviously that $\tr(\rho_i\rho_j)=0$ for $i\neq j$. 
\par Using all these facts mentioned in the above paragraph and the formula (\ref{formula}), we find that the contribution from the cross terms in eq.~(\ref{ErhoS2}) is the same
\be\label{middle} 
\mathbb{E}_U[U\rho_i U^{\dg}\otimes U\rho_j U^{\dg}]=\frac{I+\mF}{2d(d+1)}+\frac{I-\mF}{2d(d-1)}.
\ee
\par There are $d_A$ diagonal terms and $d_A^2-d_A$ cross terms. Adding all the terms together, we find
\be 
\begin{split}
&\mathbb{E}_U[\rho_S^{\otimes 2}]=\frac{I+2^{-s}\mF}{d_A(d^2-1)}-\frac{\mF+2^{-s}I}{d_Ad(d^2-1)}+\left(1-\frac{1}{d_A}\right)\left[\frac{I+\mF}{2d(d+1)}+\frac{I-\mF}{2d(d-1)}\right]\\
&=\frac{(d_Ad-2^{-s})I}{d_Ad(d^2-1)}+\frac{(2^{-s}d-d_A)\mF}{d_Ad(d^2-1)}.
\end{split}
\ee
\par The averaged purity of $\rho_B$, can be obtained from $\mathbb{E}_U[\tr(\rho_S^{\otimes2})]$ as
\be\label{ErhoB2}
\begin{split}
&\mathbb{E}_U[\tr\rho_B^2]=\mathbb{E}_U[\tr(\rho_S^{\otimes2}\mF_B\otimes I_{D})]\\
&=\frac{(d_Ad-2^{-s})d_Bd_{D}^2}{d_Ad(d^2-1)}+\frac{(2^{-s}d-d_A)d_B^2d_{D}}{d_Ad(d^2-1)}\\
&=\frac{(2^{N}-2^{-s-N_A})2^{N-N_B}}{2^{2N}-1}+\frac{(2^{N-s-N_A}-1)2^{N_B}}{2^{2N}-1},
\end{split}
\ee
where we have used the fact $d_Bd_D=d_Ad_C=d$ and denote the number of qubits in subsystem $B$ as $N_B$.
\section{Averaged purity of ``pruned" reduced state}
\par Now, we consider calculating the moments of $\rho_{B,Q}$ \textit{i.e.}, $\tr(\rho_{B,Q}^n)$. The standard strategy to compute this quantity is to use the integral representation of the projector $\Pi_q$, converting it to the charged moments of $\rho_{B,Q}$ and then evaluating the integral. This method has the advantage that in some cases it is possible to derive a formula for $\tr(\rho_{B,Q}^n)$ for general $n$. However, in our case, we focus only on the R\'enyi-2 entanglement asymmetry, as it captures the main features of entanglement asymmetry. Instead, we adopt a different method. This method was previously used in the paper \cite{Liu:2024kzv}, where the authors studied the quantum Mpemba effect in random circuit dynamics.
\par
\begin{figure}
        \centering
        \subfloat
        {\includegraphics[width=10cm]{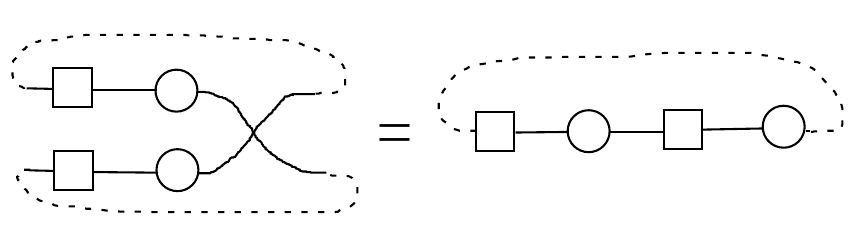}} 
        \caption{Diagram represent of identity \ref{id}.}
        \label{fig2}
\end{figure}

 Recall that the ``pruned" reduced state is defined as
\be 
\rho_{B,Q}=\sum_q\Pi_{q,B}\rho_B\Pi_{q,B}, 
\ee
where $\Pi_{q,B}$ is the projector onto the $q$-charge sector restricted on subsystem $B$. Then
\be 
\tr(\rho_{B,Q}^2)=\sum_{q=0}^{N_B}\tr(\rho_B\Pi_{B,q}\rho_B\Pi_{B,q})=\sum_{q=0}^{N_B}\tr(\rho_S^{\otimes2}(\Pi_{q,B}^{\otimes2}\mF_B)\otimes I_{D}).
\ee 
In the above equation we have used the identity
\be\label{id}
\tr(ABAB)=\tr(A^{\otimes2}(B^{\otimes2}\mF)),
\ee
which can be easily proved using the tensor network diagram. See figure \ref{fig2} for an illustration.
\par The averaged purity of $\rho_{B,Q}$ can be calculated as
\be\label{ErhoQB2} 
\begin{split}
&\mathbb{E}_U[\tr(\rho_{B,Q}^2)]=\sum_{q=0}^{N_B}\tr[\mathbb{E}_U[\rho_S^{\otimes2}](\Pi_{q,B}^{\otimes2}\mF_B)\otimes I_{D}]\\
&=\sum_{q=0}^{N_B}\tr\left[\frac{(d_Ad-2^{-s})I+(2^{-s}d-d_A)\mF}{d_Ad(d^2-1)}(\Pi_{q,B}^{\otimes2}\mF_B)\otimes I_{D}\right]\\
&=\sum_{q=0}^{N_B}\frac{(d_Ad-2^{-s})d_{D}^2}{d_Ad(d^2-1)}\tr(\Pi_{q,B}^{2})+\frac{(2^{-s}d-d_A)d_{D}}{d_Ad(d^2-1)}[\tr(\Pi_{q,B})]^{2}\\
&=\sum_{q=0}^{N_B}\frac{(d_Ad-2^{-s})d_{D}^2}{d_Ad(d^2-1)}\binom{N_B}{q}+\frac{(2^{-s}d-d_A)d_{D}}{d_Ad(d^2-1)}\binom{N_B}{q}^{2}\\
&=\frac{(d_Ad-2^{-s})d_{D}^2d_B}{d_Ad(d^2-1)}+\frac{(2^{-s}d-d_A)d_{D}}{d_Ad(d^2-1)}\binom{2N_B}{N_B}\\
&=\frac{(2^{N}-2^{-s-N_A})2^{N-N_B}}{2^{2N}-1}+\frac{(2^{N-s-N_A}-1)2^{N_B}}{2^{2N}-1}\frac{(2N_B)!}{(2^{N_B}N_B!)^2}.
\end{split}
\ee
It is important to note that all the irreducible representations of the $U(1)$ group are one-dimensional and can be labeled by an integer $q$. $\Pi_{q,B}$ is identity on the $q$-charge subspace. The dimension of the $q$-charge subspace equals to the number of occupation configurations of $q$ charges distributed on $N_B$ sites. Thus we have $\tr(\Pi_{q,B})=\tr(\Pi_{q,B}^2)=\binom{N_B}{q}$. In the above calculations, we also have used the identity
\be 
\sum_{k=0}^n\binom{n}{k}^2=\binom{2n}{n}.
\ee\\
\section{Averaged R\'enyi-2 entanglement asymmetry}
The averaged R\'enyi-2 entanglement asymmetry is defined as
\be 
\mathbb{E}_U[\D S^{(2)}(\rho_B)]=\mathbb{E}_U[\log\tr(\rho_{B}^2)]-\mathbb{E}_U[\log\tr(\rho_{B,Q}^2)].
\ee
\par To proceed, we assume that $\mathbb{E}_U[\log\tr(\rho_{B}^2)]$ and $\mathbb{E}_U[\log\tr(\rho_{B,Q}^2)]$ can be approximated by $\log\mathbb{E}_U[\tr(\rho_{B}^2)]$ and $\log\mathbb{E}_U[\tr(\rho_{B,Q}^2)]$, respectively\footnote{In fact, this approximation can be proved to be valid in the thermodynamic limit \cite{Kudler-Flam:2021rpr, Hayden:2016cfa}, which matches the regime of interest in this paper. }. From the expression of $\mathbb{E}_U[\tr(\rho_{B}^2)]$ and $\mathbb{E}_U[\tr(\rho_{Q,B}^2)]$ in eq.~(\ref{ErhoB2}) and $\ref{ErhoQB2}$ respectively, we find that if $s=N-N_A$, which means the black hole is initially maximally mixed. Then $\mathbb{E}_U[\tr(\rho_{B}^2)]=\mathbb{E}_U[\tr(\rho_{Q,B}^2)]$, which leads to $\mathbb{E}[\D S^{(2)}(\rho_B)]=0$. For $s\neq N-N_A$, the averaged R\'enyi-2 entanglement asymmetry is 
\be\label{ES2e}
\begin{split}
&\mathbb{E}[\D S^{(2)}(\rho_B)]\approx -\log\frac{\mathbb{E}_U[\tr(\rho_{B,Q}^2)]}{\mathbb{E}_U[\tr(\rho_{B}^2)]}\\
&=-\log\left[1+\frac{(2N_B)!(2^{N_B}N_B!)^{-2}-1}{\frac{2^{N+N_A}-2^{-s}}{2^{N-s}-2^{N_A}}2^{N-2N_B}+1}\right].
\end{split}
\ee
\par In the thermodynamic limit $N\rightarrow\infty, N_B\rightarrow\infty$ with $N_B/N$ kept fixed, we can use the Stirling formula $n!\approx n^ne^{-n}\sqrt{2\pi n}$ for large $n$ to simplify the factor involving factorial in the equation above as
\be 
\frac{(2N_B)!}{(2^{N_B}N_B!)^{2}}\approx \frac{1}{\sqrt{\pi N_B}}
\ee
and
\be 
\frac{2^{N+N_A}-2^{-s}}{2^{N-s}-2^{N_A}}2^{N-2N_B}=\frac{2^{N_A}-2^{-s-N}}{2^{-s}-2^{N_A-N}}2^{N-2N_B}\approx 2^{N+N_A+s-2N_B}.
\ee
Finally, in the thermodynamic limit, the averaged R\'enyi-2 entanglement asymmetry can be written as 
\be\label{ES2} 
\mathbb{E}[\D S^{(2)}(\rho_B)]= -\log\left[1+\frac{1/\sqrt{\pi N_B}-1}{2^{N+N_A+s-2N_B}+1}\right],
\ee 
which is the main result in this paper. 
\par
\begin{figure}
        \centering
        \subfloat
        {\includegraphics[width=10cm]{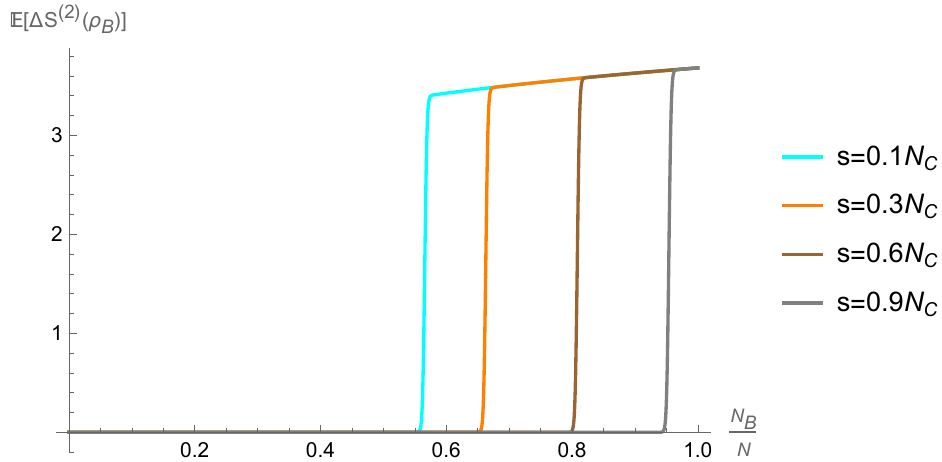}} 
        \caption{The averaged R\'enyi-2 entanglement asymmetry as a function of $\frac{N_B}{N}$ (c.f. eq.~(\ref{ES2}) or eq.~(\ref{ES2e})), where we take $N=500,N_A=15$. The maximum allowable value for $s$ is $N_C=N-N_A$.}
        \label{fig3}
\end{figure}

 From the equation above, we can see that when $N_B<(N+N_A+s)/2$, the averaged R\'enyi-2 entanglement asymmetry is almost zero. When $N_B$ begins to exceed $(N+N_A+s)/2$, the entanglement asymmetry rapidly saturates to the maximum $\frac12\log(\pi N_B)$. This behavior of the averaged R\'enyi-2 entanglement asymmetry as a function of $\frac{N_B}{N}$ is plotted in figure \ref{fig3}.
\par As shown in the figure, the averaged R\'enyi-2 entanglement asymmetry of $\rho_B$ vanishes when $N_B$ is smaller than some number that depends on the initial R\'enyi entropy $s$. When $N_B$ becomes slightly greater than that number, the entanglement asymmetry grows to its maximum sharply.
\section{Discussion and conclusion} 
Similar to the random pure state case \cite{Ares:2023ggj}, the behavior of entanglement asymmetry can be understood using the decoupling inequality. For the readers' convenience, we give the derivation in the appendix \ref{appendix}. In our case, it reads
\be\label{decineq}
\left(\mathbb{E}_U[\|\rho_B-\rho_B^{\max}\|_1]\right)^2\leq \frac{d_B}{d_D}\tr\rho_{AC}^2,
\ee
where $\rho_B^{\max}=\frac{1}{d_B}I_B$ is the maximally mixed state on $B$. $\|\mathcal{O}\|_1=\tr\sqrt{\mO^{\dg}\mO}$ is the trace norm for operator $\mO$, also called the $L_1$ norm. $\rho_{AC}$ is the initial density matrix of the $N$-qubit system:
\be
\rho_{AC}=\tr_R\rho_{ACR}=\frac{1}{d_A}I_A\otimes\rho.
\ee
Therefore we have 
\be
\tr\rho_{AC}^2=\frac{1}{d_A^2}\tr(I_A)~\tr(\rho^2)=\frac{2^{-s}}{d_A}.
\ee
Then the decoupling inequality eq.~(\ref{decineq}) becomes
\be\label{decineq1}
\left(\mathbb{E}_U[\|\rho_B-\rho_B^{\max}\|_1]\right)^2\leq \frac{1}{2^{N+N_A+s-2N_B}},
\ee
which says that if $N_B<(N+N_A+s)/2$, then $\rho_B$ becomes almost maximally mixed, while the maximally mixed state has vanishing entanglement asymmetry. Therefore we can conclude that the entanglement asymmetry of $\rho_B$ is almost vanishing for $N_B<(N+N_A+s)/2$. In particular, when the black hole is initially maximally mixed, \textit{i.e.} $s=N-N_A$, the entanglement asymmetry is always suppressed. Our result is also applicable in the late-time regime of local unitary dynamics or even Hamiltonian dynamics.
\par We know that, typically, random pure states always have the maximal possible bipartite entanglement entropies. A similar phenomenon happens here, when $N_B>(N+N_A+s)/2$, the averaged R\'enyi-2 entanglement asymmetry quickly saturates the value $\frac12\log(\pi N_B)$. Although no rigorous proofs show that this is the maximal possible value of entanglement asymmetry, there is some other evidence that this might be true. In the paper \cite{Ares:2022koq}, the authors showed that the maximal R\'enyi-$\a$ entanglement asymmetry for product states is $\frac12\log(\pi N_B\a^{1/(\a-1)}/2)$, which coincides with our result for $\a=2$. In the paper \cite{Capizzi:2023xaf}, the authors extended this result to matrix product states and to non-abelian symmetries.
\par In summary, in this paper, using the entanglement asymmetry as a modern tool to quantify symmetry breaking, we found a $U(1)$ symmetry of the radiation emerges before a certain transition time (the time when the vanishing entanglement asymmetry begins to grow) in the Hayden-Preskill thought experiment. We confirmed this emergent symmetry by calculating the entanglement asymmetry of the radiation. We found that the transition time depends on the initial entropy $s$ and the size of the diary $N_A$. When the black hole is maximally entangled with the early radiation, this emergent symmetry survives throughout the entire radiation process. Using the decoupling inequality, we can understand our findings to some extent. 
\par It's quite interesting to generalize our results to more realistic models of black hole evaporation\cite{Piroli:2020dlx} and consider the entanglement asymmetry in random unitary dynamics with structures as in \cite{Yoshida:2021xyb, Liu:2024kzv}.   
\section*{Acknowledgments}
This work was supported  by the National Natural Science Foundation of China, Grant No.\ 12465014.
\begin{appendix}
\section{The decoupling inequality}\label{appendix}
In this section, we give the derivation of eq.~\ref{decineq}. This is a simplified version of the so-called decoupling inequality. However, it is enough for our purpose. For the more general case, we encourage the reader to consult the reference \cite{Preskill:1998}. For the readers' convenience, we restate the inequality here:
\be\label{decineq2} 
\left(\mathbb{E}_U[\|\rho_B-\rho_B^{\max}\|_1]\right)^2\leq \frac{d_B}{d_D}\tr\rho_{AC}^2.
\ee
\par To prove it, we also need to introduce the $L_2$ norm of an operator $\mO$, which is defined as $\|\mO\|_2=\sqrt{\tr(\mO^{\dg}\mO)}$. It's easy to show that the $L_1$ norm and $L_2$ norm satisfy
\be\label{normineq} 
\|\mO\|_1\leq\sqrt{d_{\mO}}\|\mO\|_2,
\ee
where $d_{\mO}$ is the dimension of the Hilbert space that $\mO$ acts on.
\par By Jensen's inequality, we know that for any non-negative function $f$, one has
\be\label{Jensen} 
(\mathbb{E}_U[f(U)])^2\leq\mathbb{E}_U[f(U)^2].
\ee
A combination of eq.~(\ref{normineq}) and eq.~(\ref{Jensen}) gives
\be\label{A4} 
\left(\mathbb{E}_U[\|\rho_B-\rho_B^{\max}\|_1]\right)^2\leq\mathbb{E}_U[\|\rho_B-\rho_B^{\max}\|_1^2]\leq d_B\mathbb{E}_U[\|\rho_B-\rho_B^{\max}\|_2^2].
\ee
The right-hand-side of eq.~(\ref{A4}) reminds us to compute
\be 
\|\rho_B-\rho_B^{\max}\|_2^2=\tr(\rho_B-\rho_B^{\max})^2=\tr(\rho_B^2)-\frac{1}{d_B},
\ee
and
\be 
\mathbb{E}_U[\tr\rho_B^2]=\mathbb{E}_U[\tr(\rho_S^{\otimes2}\mF_B\otimes I_{D})].
\ee
Substituting $\rho_S=U\rho_{AC}U^{\dg}$ into the equation above, we obtain
\be\label{ErhoB} 
\mathbb{E}_U[\tr\rho_B^2]=\mathbb{E}_U[\tr(U^{\dg\otimes2}(\mF_B\otimes I_{D})U^{\otimes2}\rho_{AC}^{\otimes2})], 
\ee
where we have used the cyclicity of the trace.
\par Now use the identity
\be\label{identity} 
\begin{split}
\mathbb{E}_U[U^{\dg\otimes2}(\mF_B\otimes I_{D})U^{\otimes2}]=\frac{1}{d_B}\frac{1-1/d_D}{1-1/d^2}I+\frac{1}{d_D}\frac{1-1/d_B}{1-1/d^2}\mF.
\end{split}
\ee
Here, as before, we have used the notation that operators without subscripts such as $U,I$ and $\mF$ act on the full system $S$.
This identity is somewhat complicated to prove. In order not to affect the coherence of the main proof, we will give the proof at the end of this section.
\par Since $d=d_Bd_D$, which means $d>d_B$ and $d>d_D$. Additionally, from eq.~(\ref{identity}), we know that
\be 
\mathbb{E}_U[U^{\dg\otimes2}(\mF_B\otimes I_{D})U^{\otimes2}]\leq\frac{1}{d_B}I+\frac{1}{d_D}\mF.
\ee
Using the inequality above and substituting it into equation \ref{ErhoB}, we find
\be 
\begin{split}
&\mathbb{E}_U[\tr\rho_B^2]\leq \tr\left(\left(\frac{1}{d_B}I+\frac{1}{d_D}\mF\right)\rho_{AC}^{\otimes2}\right)\\
&=\frac{1}{d_B}(\tr\rho_{AC})^2+\frac{1}{d_D}\tr(\rho_{AC}^2)=\frac{1}{d_B}+\frac{1}{d_D}\tr(\rho_{AC}^2).
\end{split}
\ee
Finally, we have
\be 
d_B\mathbb{E}_U[\|\rho_B-\rho_B^{\max}\|_2^2]\leq\frac{d_B}{d_D}\tr(\rho_{AC}^2).
\ee
From eq.~(\ref{A4}) we know that the decoupling inequality eq.~(\ref{decineq2}) is indeed valid.
\par Now we give the proof of the identity eq.~(\ref{identity}). It's useful to introduce the following two types of maximally entangled states:
\be 
\begin{split}
&\ket{\mathcal{I^+}}=\ket{\mathcal{I^+}}_B\ket{\mathcal{I^+}}_D=\sum_{\a\b\g\d}\ket{\a\a\b\b}_B\ket{\g\g\d\d}_D,\\ &\ket{\mathcal{I^-}}=\ket{\mathcal{I^-}}_B\ket{\mathcal{I^-}}_D=\sum_{\a\b\g\d}\ket{\a\b\a\b}_B\ket{\g\d\g\d}_D,
\end{split}
\ee
where the summation runs from $\a,\b=1,2,\cdots,d_B$, $\g,\d=1,2,\cdots,d_D$. The Choi-Jamiolkowski mapping allows us to view operators on the two ``replicas" as states:
\be 
\ket{\mO\otimes\mO}=(\mathbb{I}\otimes\mO\otimes\mathbb{I}\otimes\mO)\ket{\mI^{+}}.
\ee
In this way, the identity operator $I$ and swap operator $\mF$ are mapped to $\ket{I}=\ket{\mI^+}$ and $\ket{\mF}=\ket{\mI^-}$, respectively. Then by the Choi-Jamiolkowski isomorphism, the identity eq.~(\ref{identity}) is equivalent to 
\be\label{identity1} 
\bra{\mI_+}_B\bra{\mI_-}_D\mathbb{E}_U[U\otimes U^*\otimes U\otimes U^*]=\frac{1}{d_B}\frac{1-1/d_D}{1-1/d^2}\bra{\mI_+}+\frac{1}{d_D}\frac{1-1/d_B}{1-1/d^2}\bra{\mI_-}.
\ee
From the Haar identity for the fourth moment of $U$
\be 
\mathbb{E}_U[U\otimes U^*\otimes U\otimes U^*]=\frac{1}{d^2-1}\left[(\ket{\mI_+}\bra{\mI_+}+\ket{\mI_-}\bra{\mI_-})-\frac{1}{d}(\ket{\mI_+}\bra{\mI_-}+\ket{\mI_-}\bra{\mI_+})\right],
\ee
and the facts that $\braket{\mI_{\pm}}{\mI_{\pm}}=d^2$ and $\braket{\mI_{\pm}}{\mI_{\mp}}=d$ (similar expression for subsystem $B,D$), we can easily check that eq.~(\ref{identity1}) is indeed satisfied. In fact, the identity eq.~(\ref{formula}) can also be easily proved using this operator-state mapping method.
\end{appendix}
\bibliography{2024}
\bibliographystyle{ieeetr}
\end{document}